\documentclass[aps,prl,twocolumn,showpacs,superscriptaddress,floatfix,10pt]{revtex4-1}

\usepackage{graphicx}
\usepackage{amsmath}
\usepackage{epsfig}
\usepackage{helvet}
\usepackage{amssymb}

\newcommand{\bk}[1]{\mbox{$\langle #1 \rangle$}}
\newcommand{\tTr}[1]{\mbox{$\text{tTr}\left( #1 \right)$}}
\newcommand{\tr}{\mbox{$\text{tr}$}}

\def\CDL{\mbox{\scriptsize CDL}}
\def\TRI{\mbox{\scriptsize triv}}
\def\CRI{\mbox{\scriptsize crit}}
\begin{document}

\title{Tensor Network Renormalization}

\author{G. Evenbly}
\affiliation{Institute for Quantum Information and Matter, California Institute of Technology, Pasadena CA 91125, USA}
\email{evenbly@caltech.edu}
\author{G. Vidal}
\affiliation{Perimeter Institute for Theoretical Physics, Waterloo, Ontario N2L 2Y5, Canada}  \email{gvidal@perimeterinstitute.ca}
\date{\today}

\begin{abstract}
We introduce a coarse-graining transformation for tensor networks that can be applied to study both the partition function of a classical statistical system and the Euclidean path integral of a quantum many-body system. The scheme is based upon the insertion of optimized unitary and isometric tensors (\textit{disentanglers} and \textit{isometries}) into the tensor network and has, as its key feature, the ability to remove short-range entanglement/correlations at each coarse-graining step. Removal of short-range entanglement results in scale invariance being explicitly recovered at criticality. In this way we obtain a proper renormalization group flow (in the space of tensors), one that in particular (i) is computationally sustainable, even for critical systems, and (ii) has the correct structure of fixed points, both at criticality and away from it. We demonstrate the proposed approach in the context of the 2D classical Ising model.  
\end{abstract}


\maketitle

Understanding emergent phenomena in many-body systems remains one of the major challenges of modern physics. With sufficient knowledge of the microscopic degrees of freedom and their interactions, we can write the \textit{partition function} of a classical system, namely a weighted sum of all its microscopic configurations; or the analogous \textit{Euclidean path integral} of a quantum many-body system, where the weighted sum is now over all conceivable trajectories. These objects contain complete information on the collective properties of the many-body system. However, evaluating partition functions or Euclidean path integrals is generically very hard. Kadanoff's spin-blocking procedure \cite{Kadanoff} opened the path to non-perturbative approaches based on coarse-graining a lattice \cite{Wilson,DMRG}. More recently, Levin and Nave proposed the tensor renormalization group (TRG) \cite{TRG}, a versatile real-space coarse-graining transformations for 2D classical partition functions ---or, equivalently, Euclidean path integrals of 1D quantum systems.   

TRG is an extremely useful approach that has revolutionized how we coarse-grain lattice models \cite{TRG,TRGenv,SRG, TEFR, TRGplus, HOTRG,TRG3D}. However, this method fails to remove part of the short-range correlations in the partition function and, as a result, the coarse-grained system still contains irrelevant microscopic information. Conceptually, this is in conflict with the very spirit of the renormalization group (RG) and results in an RG flow with the wrong structure of non-critical fixed points, as discussed in Ref. \cite{TEFR}. Computationally, the accumulation of short-range correlations over successive TRG coarse-graining transformation also has important consequences: as pointed out by Levin and Nave, it implies the \textit{breakdown} of TRG at criticality \cite{TRG, breakdown} (although universal information, such as critical exponents, can still be obtained from finite systems).

An analogous problem, faced earlier in the context of ground state wave-functions, was resolved with the introduction of entanglement renormalization techniques \cite{ER,Algorithms}. In this \textit{Letter} we will adapt those techniques to the coarse-graining of partition functions/Euclidean path integrals, and will demonstrate that the resulting scheme generates a proper RG flow, with the correct structure of critical and non-critical fixed points. 

A distinctive feature of our proposal, which we call \textit{tensor network renormalization} (TNR), is that it removes short-range correlations from the partition function at \textit{each} coarse-graining step. In this way, the effective tensor network description at a given length scale is free from irrelevant microscopic details belonging to shorter length scales. The upshot is a computationally sustainable RG flow in the space of tensors. At criticality, removal of short-range correlations circumvents TRG's breakdown, while scale invariance is explicitly realized in the form of a critical fixed-point tensor \cite{fixedpoint}. This allows us to effectively address infinitely large systems, avoiding finite-size effects. Off-criticality we also obtain the correct structure of non-critical RG fixed points. In this case, the corresponding fixed-point tensors coincide with those previously obtained by Gu and Wen's tensor-entanglement-filtering renormalization (TEFR) \cite{TEFR, filterCDL}. For simplicity, we demonstrate TNR with the statistical partition function of the 2D classical Ising model (equivalently, the Euclidean path integral of the 1D quantum Ising model with transverse magnetic field), although the key ideas apply to general two dimensional lattice models and extend to higher dimensions.

\begin{figure}[!t]
\begin{center}
\includegraphics[width=7cm]{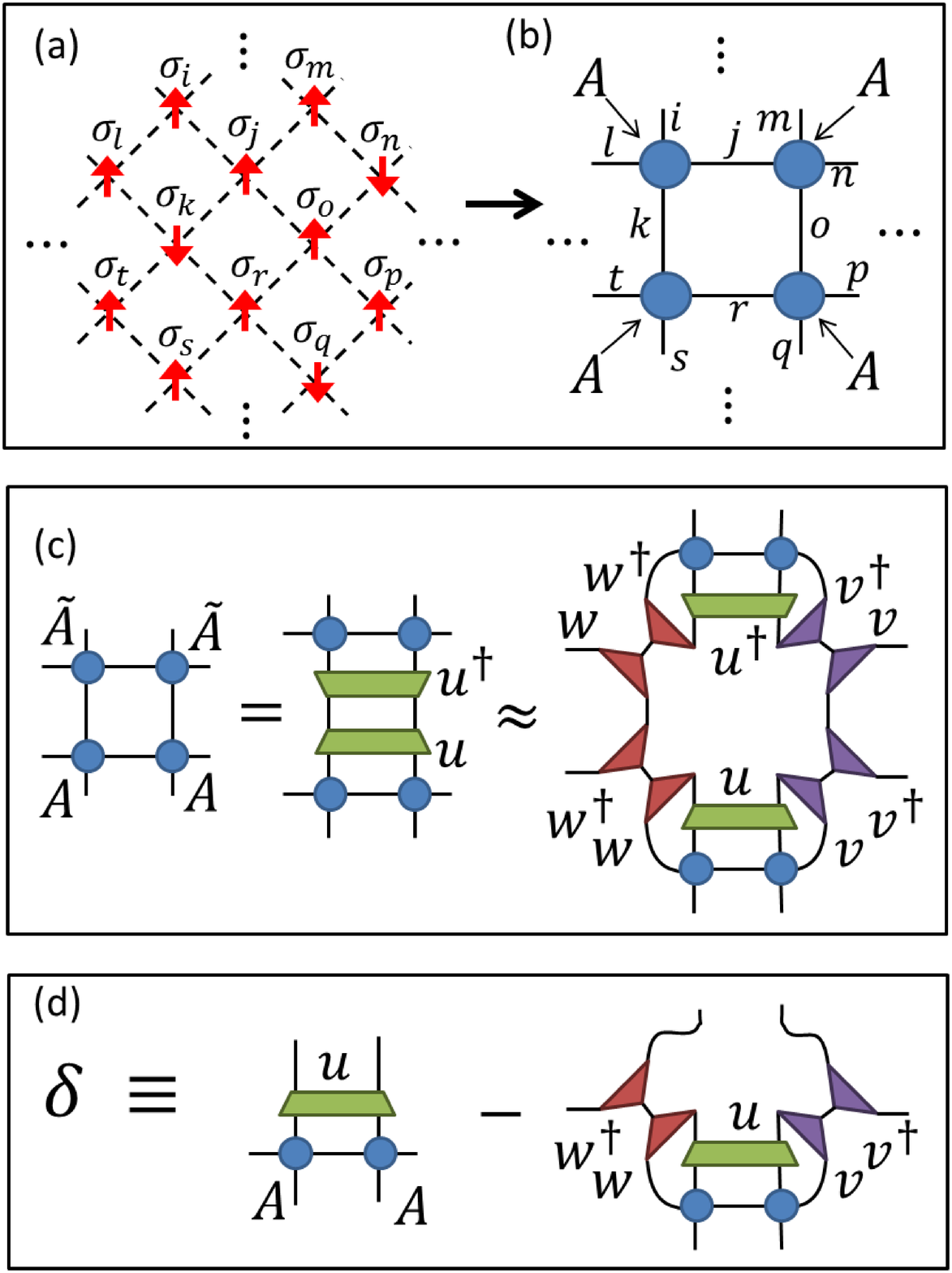}
\caption{(a) As an example, we consider a square lattice (slanted $45^{\circ}$) of classical spins, where $\sigma_k \in \{+1,-1\}$ is an Ising spin on site $k$. (b) Graphical representation of a part of the tensor network, where each circle denotes a tensor $A$, for the partition function $Z$ of the classical spin model, see Eq. \ref{eq:Z}. Here tensor $A_{ijkl}$ encodes the Boltzmann weights of the spins $\{\sigma_i,\sigma_j,\sigma_k,\sigma_l \}$ according to the Hamiltonian function $H$, see Eq. \ref{eq:A}. (c) Insertion of a pair of disentanglers $uu^{\dagger}$ between four tensors, where tensors $\tilde A$ are obtained from tensors $A$ through a gauge transformation on their horizontal indices \cite{GaugeChange}, followed by an insertion of four projectors of the form $vv^{\dagger}$ (or $ww^{\dagger}$). These projectors  introduce a truncation error. (d) Tensor $\delta$, whose norm $\| \delta \|$ measures the truncation error introduced by the isometries $v$ and $w$. Disentanglers and isometries are chosen so as to minimize $\| \delta \|$.}
\label{fig:uvw}
\end{center}
\end{figure}

\textit{Renormalization group flow in the space of tensors.---} Our starting point, also used in TRG \cite{TRG,TRGenv,SRG,TEFR,TRGplus,HOTRG,TRG3D}, is the observation that the partition function $Z$ of a 2D classical system (for concreteness assumed to be translation invariant on a periodic square lattice) can be re-written as a 2D tensor network made of $N$ copies of some tensor A (see Eq. \ref{eq:A} and Fig. \ref{fig:uvw}(a,b) for an explicit example), 
\begin{equation} \label{eq:Z}
Z = \sum_{ijk\cdots} A_{ijkl}  A_{mnoj}  A_{krst}  A_{opqr} \cdots \equiv \tTr{\otimes_{x=1}^{N} A}.
\end{equation} 
Here each index hosts a $\chi$-level local degree of freedom (e.g. $i=1,2,\cdots, \chi$), the tensor components $A_{ijkl}$ are local weights, and the tensor trace tTr denotes a sum over configurations of all the indices $ijk\cdots$. 

Our goal is to produce an effective tensor $A^{(1)}$, roughly accounting for four copies of the original tensor $A^{(0)} \equiv A$, such that $Z$ can be approximately expressed as a coarser tensor network made of just $N/4$ copies of $A^{(1)}$, $Z \approx \tTr{\otimes_{x=1}^{N/4} A^{(1)}}$. By iteration, a sequence of tensors 
\begin{equation} \label{eq:flow}
A^{(0)} \rightarrow A^{(1)} \rightarrow  A^{(2)} \rightarrow \cdots
\end{equation}
will be produced such that, for any length scale $s$,
\begin{equation}
Z \approx \tTr{\otimes_{x=1}^{N_s} A^{(s)}},~~~~~~N_s \equiv N/4^s.
\end{equation}
Thus, after $\tilde{s} \equiv \log_4 (N)$ iterations [assuming $N = 4^{\tilde{s}}$ for some integer $\tilde{s}>0$], the partition function $Z$ becomes the trace of a single tensor $A^{ ( \tilde{s} )}$, $Z \approx \sum_{ij} A^{(\tilde{s})}_{ijij}$, which we can finally evaluate \cite{npoint}. On the other hand, in the thermodynamic limit $N\rightarrow \infty$, we can study the flow in the space of tensors given by Eq. \ref{eq:flow}. The fixed-point tensors of this flow will capture the universal properties of the phases and phase transitions of the system.

\begin{figure}[!b]
\begin{center}
\includegraphics[width=7cm]{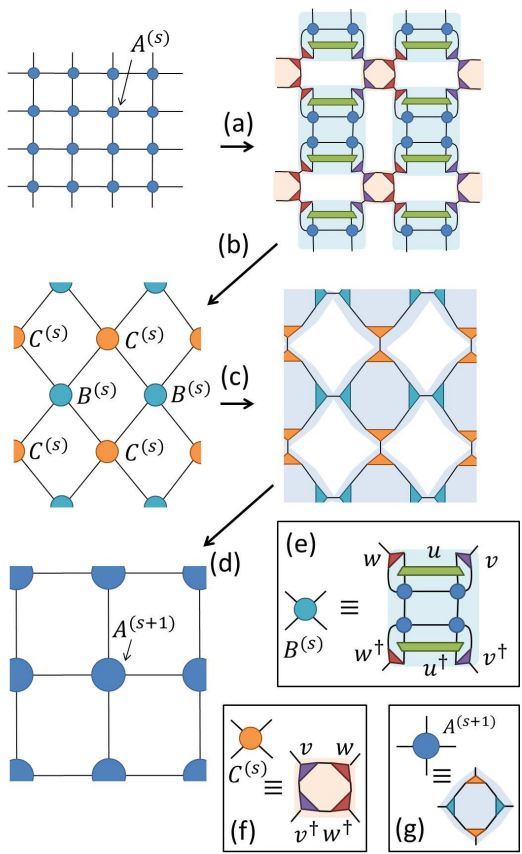}
\caption{
Steps (a)-(d) of a TNR transformation to produce tensor $A^{(s+1)}$ from tensor $A^{(s)}$. In step (a), the insertion of disentanglers $u$ and isometries $v$ and $w$ is made according to Fig. \ref{fig:uvw}(b). Insets (e)-(g) contain the definition of the auxiliary tensors $B^{(s)}$ and $C^{(s)}$ and the coarse-grained tensor $A^{(s+1)}$.
}
\label{fig:TNR}
\end{center}
\end{figure}

\textit{Tensor Network Renormalization.---} Our coarse-graining transformation for the partition function $Z$ in Eq. \ref{eq:Z} is based on  locally inserting (exact or approximate) resolutions of the identity into the tensor network. The goal is to reorganize the local degrees of freedom, so as to be able to identify and remove those that are only correlated at short distances. 

Let us regard each index of the network as hosting a $\chi$-dimensional complex vector space $\mathbb{V} \equiv \mathbb{C}^{\chi}$. We consider two types of insertions, see Fig. \ref{fig:uvw}(c). The first type is implemented by a pair $uu^{\dagger}=I^{\otimes 2}$ of unitary transformations $u$, or \textit{disentanglers}, acting on two neighboring indices,  $u: \mathbb{V}\otimes \mathbb{V} \rightarrow \mathbb{V} \otimes \mathbb{V}$. The disentanglers $u$ will be used to remove short-range correlations \cite{disentanglers}.  
Notice that inserting a pair of disentanglers $uu^{\dagger}$ does not change the partition function $Z$ represented by the tensor network. 

The second type of insertion is implemented by a projector $vv^{\dagger}$ (or $ww^{\dagger}$), where the \textit{isometry} $v$ (or $w$) combines two indices into a single one, $v: \mathbb{V} \rightarrow \mathbb{V} \otimes \mathbb{V}$, with $v^{\dagger}v = I$. Since $vv^{\dagger}$ is not the identity but a $\chi$-dimensional projector acting on the $\chi^2$-dimensional space $\mathbb{V}\otimes \mathbb{V}$, inserting it into the tensor network introduces a \textit{truncation error} into the representation of the partition function $Z$. This error can be estimated by the norm $\| \delta \|$ of the difference operator $\delta$ defined in Fig. \ref{fig:uvw}(d). If, somehow, only a small truncation error $\| \delta \|$ is introduced, then the resulting tensor network will still be a good approximation to the partition function $Z$. 

Fig. \ref{fig:TNR} shows graphically the proposed TNR transformation. 
In step (a), disentanglers and isometries are inserted between blocks of $2\times 2$ tensors $A^{(s)}$. 
In step (b) two types of auxiliary tensors, $B^{(s)}$ and $C^{(s)}$, are produced by contracting indices. 
In step (c) tensors $B$ and $C$ are split using a singular value decomposition, as it is done in TRG \cite{WeightC}. 
Finally, in step (d) the coarse-grained tensor $A^{(s+1)}$ at scale $s+1$ is obtained by further contraction of indices. 
The disentanglers and isometries introduced in step $(a)$ are chosen so as to minimize the truncation error $\| \delta \|$ in Fig. \ref{fig:uvw}(d), using well-established, iterative optimization methods for unitary and isometric tensors \cite{Algorithms}, which are further detailed in Ref.\cite{TNRalgorithms}. The overall computational cost of computing tensor $A^{(s+1)}$ from tensor $A^{(s)}$ scales as $O(\chi^{7})$, although this cost can be reduced to $O(\chi^6)$ through introducing controlled approximations \cite{TNRalgorithms}.
 
To gain some insight into how TNR operates, let us consider first an oversimplified scenario where the partition function $Z$ \textit{only} contains short-range correlations (technically, this corresponds to a so-called CDL tensor \cite{AppendixB}). If we set the disentanglers to be trivial, $u = I^{\otimes 2}$, then the coarse-graining transformation reduces to TRG and fails to remove the short-range correlations. However, with a judicious choice of disentanglers $u$ these correlations are removed and an uncorrelated, trivial tensor $A^{\TRI}$ is produced \cite{AppendixB, TEFRalso}. Therefore the role of  disentanglers is to remove short-range correlations. Their action will be particularly important at criticality, where correlations are present at all length scales. 

\begin{figure}[!tbph]
\begin{center}
\includegraphics[width=8cm]{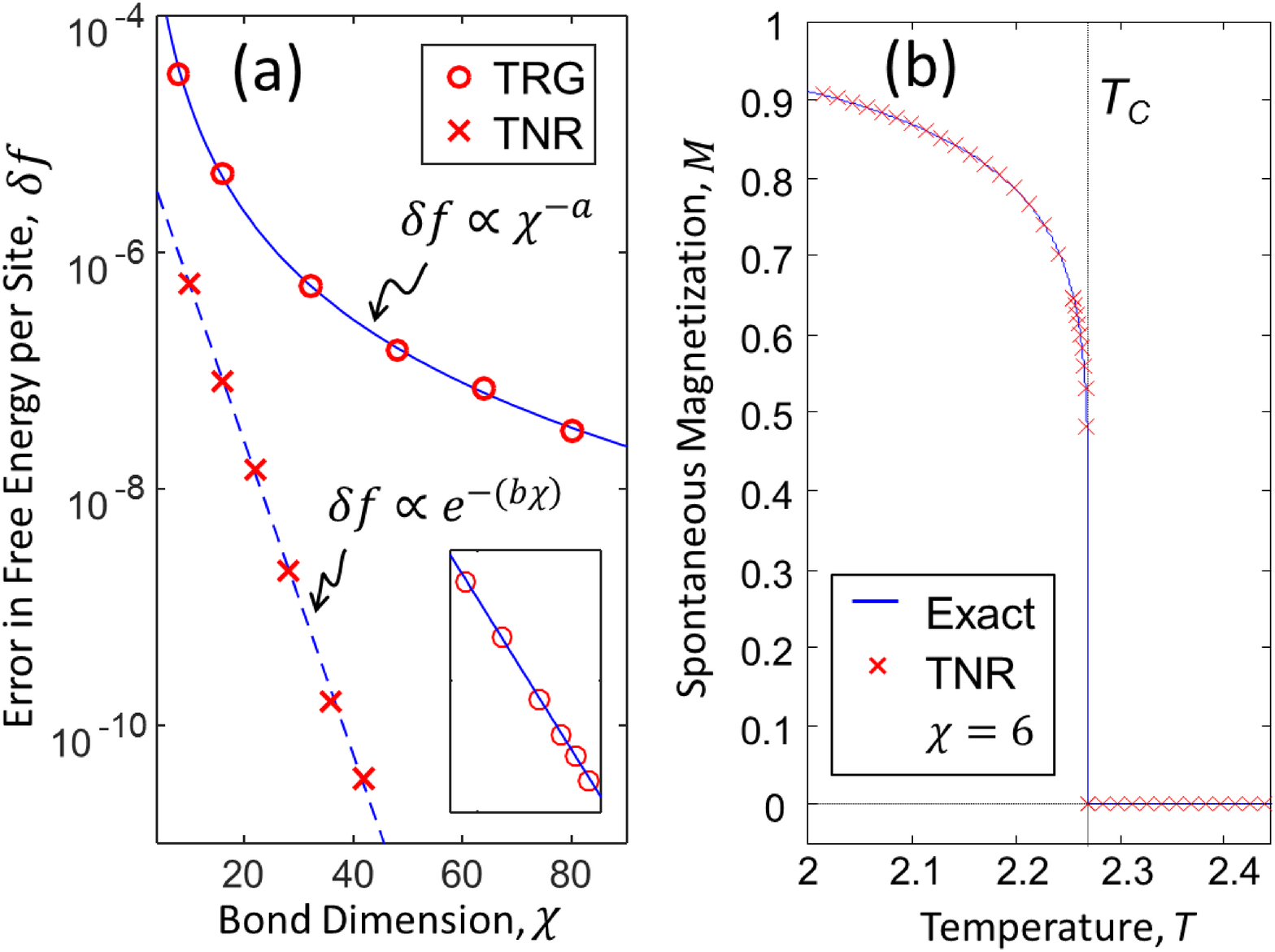}
\caption{Benchmark results for the square lattice Ising model on a lattice with $2^{39}$ spins. (a) Relative error in the free energy per site ${\delta}{f}$ at the critical temperature $T_c$, comparing TRG and TNR over a range of bond dimensions $\chi$. The TRG errors fit $\delta f \propto \chi^{-3.02}$ (the inset displays them using log-log axes), while TNR errors fit $\delta f \propto \exp(-0.305 \chi)$. Extrapolation suggests that TRG would require bond dimension $\chi \approx 750$ to match the accuracy of the $\chi = 42$ TNR result. (b) Spontaneous magnetization $M(T)$ obtained with TNR with $\chi=6$ \cite{fixedboundary}. Even very close to the critical temperature, $T = 0.9994 \ T_c$, the magnetization $M\approx 0.48$ is reproduced to within $1\%$ accuracy.}
\label{fig:EnNum}
\end{center}
\end{figure}

\begin{figure}[!tbph]
\begin{center}
\includegraphics[width=8.5cm]{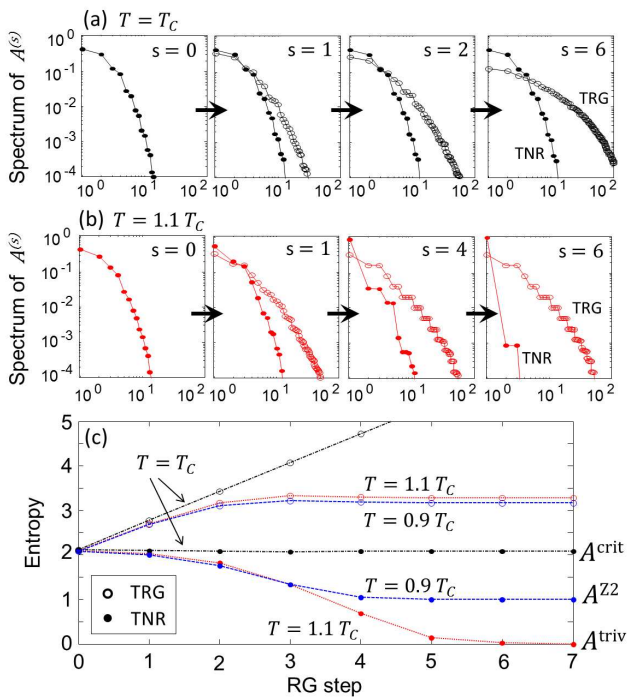}
\caption{(a) Singular values $\lambda_{\alpha}$ of the matrix $[A^{(s)}]_{(ij)(kl)}$ obtained after $s$ RG steps \cite{RGstep} using TNR (filled circles) or TRG (empty circles) for the 2D Ising model at critical temperature $T_C$. (b) Singular values for $T=1.1 \ T_C$. (c) Plot of the Von-Neumann entropy $-\sum_{\alpha} \tilde{\lambda}_{\alpha}\log(\tilde{\lambda}_{\alpha})$ of the (normalized) singular values of tensors $[A^{(s)}]_{(ij)(kl)}$ obtained with TRG (empty circles) or TNR (filled circles).}
\label{fig:FlowSpectrum}
\end{center}
\end{figure}

\textit{Example: Partition function of the 2D classical Ising model.---} We consider the partition function
\begin{equation} \label{eq:Ising}
Z = \sum_{\{\sigma\}} e^{-H(\{\sigma\})/T}, ~~~H\left( \{ \sigma \} \right) =  - \sum_{\left\langle i,j \right\rangle}  \sigma _i\sigma _j
\end{equation}
on the square lattice, where $\sigma_k \in \{+1,-1\}$ is an Ising spin on site $k$ and $T$ denotes the temperature. Recall that this model has a global $Z_2$ symmetry: it is invariant under the simultaneous flip $\sigma_k \rightarrow -\sigma_k$ of all the spins. 
We obtained an exact representation for the tensor $A$ in Eq. \ref{eq:Z} in terms of four Boltzmann weights $e^{\sigma_i\sigma_j/T}$,
\begin{equation} \label{eq:A}
A_{ijkl} = e^{\left(\sigma_i\sigma_j + \sigma_j\sigma_k + \sigma_k\sigma_l + \sigma_l\sigma_i \right)/T},
\end{equation}
which corresponds to having one tensor $A$ for every two spins, and a tensor network with a $45$ degree tilt with respect to the spin lattice, see Fig. \ref{fig:uvw}(a,b). We actually built our starting tensor $A^{(0)}$, with bond dimension $\chi=4$, by joining a square block of four tensors $A$ together. We then applied up to 18 TNR transformations to a system made of $N = 2^{18} \times 2^{18}$ tensors $A^{(0)}$, or equivalently $2\times 4 \times N$ Ising spins.

Fig. \ref{fig:EnNum}(a) shows the relative error $\delta f$ in the free energy per site $f \equiv \log(Z)/(8N)$, at the critical temperature $T_c \equiv 2/\ln(1+\sqrt{2}) \approx 2.269$, for both TRG and TNR as a function of the bond dimensions $\chi$ \cite{chiexplain}. The TRG error decays polynomially, while the TNR error is reduced exponentially, showing a qualitatively different behaviour and implying that significantly more accurate results can be obtained with TNR. Figure \ref{fig:EnNum}(b) shows the spontaneous magnetization $M(T)$ as a function of temperature $T$ obtained with TNR for $\chi=6$. Even close to $T=T_c$, we see remarkable agreement with the exact solution. 
 
However, the most significant feature of TNR is revealed in Fig. \ref{fig:FlowSpectrum}, which shows, as a function of the scale $s$, the spectrum of singular values of tensor $A^{(s)}$ when regarded as a matrix $[A^{(s)}]_{(ij)(kl)}$. 
Fig. \ref{fig:FlowSpectrum}(a) considers the critical point, $T=T_c$, and shows that under TNR, the spectrum of $A^{(s)}$ quickly becomes independent of the scale $s$. This has two major implications. On the one hand, it is a strong evidence that $A^{(s)}$ itself has converged to a fixed-point tensor $A^{\CRI}$ (up to small corrections, see \cite{AppendixA} for detials), thus recovering the characteristic scale invariance expected at criticality. 
On the other hand, it implies that the bond dimension $\chi$ required to maintain a fixed, small truncation error $\| \delta \|$ is essentially independent of scale, and thus so is also the computational cost. That is, we have obtained a computationally sustainable RG transformation.
In sharp contrast, the spectrum generated by TRG develops a growing number of large singular values as we increase the scale $s$, indicating that the tensor is not scale invariant. The bond dimension (and thus the computational cost at constant truncation error) grows rapidly with scale. This growth is caused by the accumulation of short-range correlations at each iteration and pinpoints the breakdown of TRG at a critical point \cite{AppendixC}. 

Fig. \ref{fig:FlowSpectrum}(b) considers a slightly larger temperature, $T=1.1 \ T_c$. Now TNR generates a flow towards the trivial fixed-point tensor $A^{\TRI}$, characterized by just one non-zero singular value, which represents the infinite temperature, disordered phase. As expected of a proper RG scheme, for any $T> T_c$ the flow is to the same trivial fixed-point tensor $A^{\TRI}$. In contrast, for any $T>T_c$, TRG generates a flows to a fixed-point tensor that depends on the initial temperature $T$. In other words, failure to remove some of the short-range correlations implies that (RG irrelevant) microscopic information has been retained during coarse-graining, contrary to the spirit of the RG flow. For $T<T_c$ (not shown) we obtain a similar picture as for $T>T_c$. However, now each eigenvalue in the spectrum has degeneracy 2, and TNR flows to a new fixed-point tensor $A^{Z_2} \equiv A^{\TRI}\oplus A^{\TRI}$ corresponding to the $Z_2$ symmetry breaking, ordered phase ---the $Z_2$ spin flip symmetry acts on $A^{Z_2}$ by exchanging its two copies of $A^{\TRI}$. Finally, Fig. \ref{fig:FlowSpectrum}(c) uses the entropy of the spectrum as a function of scale to visualize the RG flow towards one of the three fixed-point tensors: the ordered $A^{Z_2}$ for $T<T_c$, the critical $A^{\CRI}$ for $T=T_c$, and the disordered $A^{\TRI}$ for $T>T_c$. 

\textit{Outlook.---} We have proposed \textit{tensor network renormalization} (TNR), a coarse-graining transformation for tensor networks that produces a proper RG flow in the space of tensors, and demonstrated its performance for $2D$ classical partition functions --including the explicit recovery of scale invariance at the critical point. When applied to the Euclidean path integral of the 1D quantum Ising model (after suitable discretization in the imaginary time direction), it produced results similar to the ones described above. The approach can also be used to compute the norm $\bk{\Psi | \Psi}$ of a 2D quantum many-body state encoded in a PEPS \cite{PEPS}.

TNR borrows its key idea ---the use of disentanglers--- from entanglement renormalization \cite{ER,Algorithms}, the coarse-graining scheme for many-body wave-functions that led to the multi-scale entanglement renormalization ansatz (MERA) \cite{MERA,MERACFT}. The two approaches turn out to be deeply connected: when applied to the Euclidean path integral of a Hamiltonian $H$, TNR produces a MERA for the ground and thermal states of $H$ \cite{TNRtoMERA}. 
     
We thank Z.-C. Gu and X.-G. Wen for clarifying aspects of their TEFR approach \cite{TEFR}. G.E. is supported by the Sherman Fairchild Foundation. G. V. acknowledges support by the John Templeton Foundation and NSERC. The authors also acknowledge support by the Simons Foundation (Many Electron Collaboration). Research at Perimeter Institute is supported by the Government of Canada through Industry Canada and by the Province of Ontario through the Ministry of Research and Innovation.


\newpage
\appendix
\section{Appendix A.-- RG flow in the space of tensors: the $2D$ classical Ising model.} 

In this appendix we provide additional details on the flow that TNR generates when applied to a tensor network representation of the partition function of the 2D classical Ising model on a square lattice, as defined in the main text.

For all possible values of the temperature $T$ in the Ising model, the flow is seen to always end in one of three possible fixed-point tensors: the ordered $A^{Z_2}$ for $T<T_c$, the critical $A^{\CRI}$ for $T=T_c$, and the disordered $A^{\TRI}$ for $T>T_c$. The ordered and disorder fixed-point tensors can be expressed exactly with a finite bond dimension, namely $\chi=2$ and $\chi=1$, respectively, whereas an exact expression of the critical fixed-point tensor $A^{\CRI}$ is suspected to require an infinite bond dimension, and thus here we can only obtain an approximate representation. We emphasize that the non-critical fixed-point tensors $A^{Z_2}$ and $A^{\TRI}$ are equivalent to those previously obtained by TEFR \cite{TEFR}.

For purposes of clarity, instead of following the flow of tensors $A^{(s)}$ we will display instead the flow of the auxiliary tensors $B$ that appear in an intermediate step of the coarse-graining transformation, see Fig.\ref{fig:TNR} in the main text, as these tensors have smaller bond dimension \cite{chiexplain}, that is less coefficients, and are thus more easily plotted. However, their behavior under the RG flow is seen to be essentially equivalent to that of tensors $A$.

\begin{figure}[!tbph]
\begin{center}
\includegraphics[width=8cm]{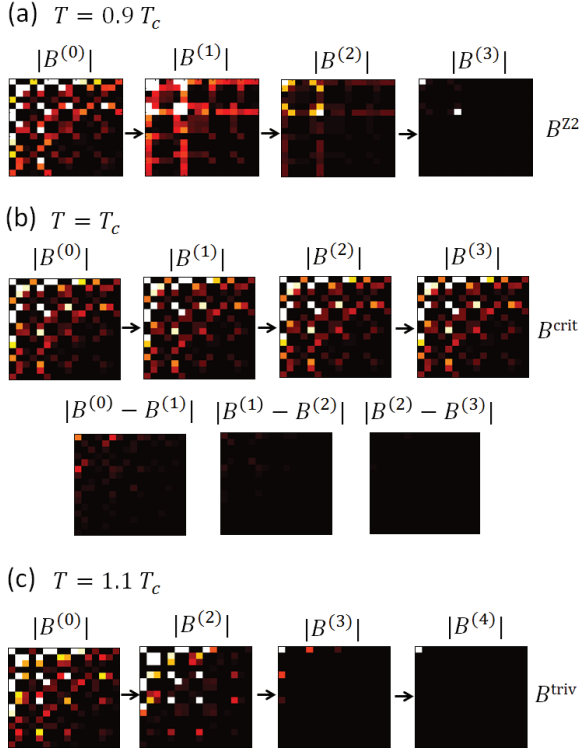}
\caption{Plots of the elements of tensors $[B^{(s)}]_{ijkl}$, when reshaped as $16\times 16$ matrices, after $s$ iterations of the TNR coarse-graining transformation, for several values of $s$. Dark pixels indicate elements of small magnitude and lighter pixels indicate elements with larger magnitude. (a) Starting at a sub-critical temperature, $T = 0.9\  T_C$, the coarse-grained tensors quickly converge to the $Z_2$ fixed-point tensor $B^{Z_2}\equiv B^{\TRI}\oplus B^{\TRI}$. (b) Starting at the critical temperature, $T = T_C$, the coarse-grained tensors converge to a non-trivial fixed-point tensor $B^{\CRI}$. Notice that the difference between coarse-grained tensors, $| {{B^{(s)}} - {B^{(s + 1)}}} |$, which is displayed with the same color intensity as the plots of $| B^{(s)} |$, is already very small (as compared to the magnitude of the elements in the individual tensors) for $s=1$. (c) Starting at the super-critical temperature, $T = 1.1\  T_C$, the coarse-grained tensors quickly converge to the disordered fixed point $B^{\TRI}$, that has only one non-zero element.}
\label{fig:RGflowimage}
\end{center}
\end{figure}

Starting with the partition function $Z$ of the $2D$ classical Ising model at a temperature $T$, we coarse grain the corresponding tensor network iteratively using TNR. This generates a sequence of tensors $B^{(s)}$, for multiple RG steps $s=0,1,2,\ldots$. Let $[B^{(s)}]_{ijkl}$ denote the components of $B^{(s)}$. Here we consider the case where each index ($i$, $j$, $k$, and $l$) has dimension $4$.  The elements of these tensors, reshaped as $16\times 16$ matrices $[B^{(s)}]_{(ij)(kl)}$ and then normalized such that their singular values sum to unity, are plotted in Fig. \ref{fig:RGflowimage}.

\begin{figure}[!tbph]
\begin{center}
\includegraphics[width=7cm]{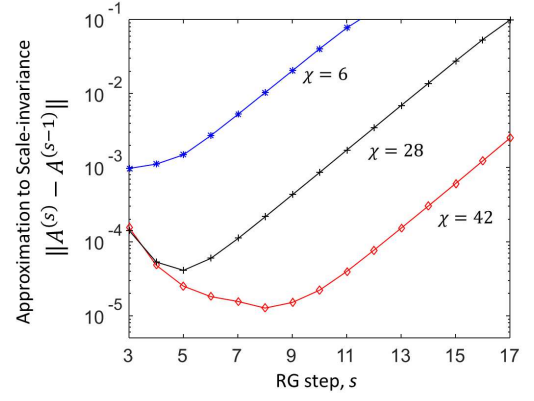}
\caption{The precision with which TNR approximates a scale-invariant fixed point tensor for the $2D$ classical Ising model at critical temperature $T_c$ is examined by comparing the difference between tensors produced by successive TNR iterations ${\delta ^{(s)}} \equiv \| {{A^{(s)}} - {A^{(s - 1)}}} \|$, where tensors have been normalized such that $\| {{A^{(s)}}} \| = 1$. For small $s$ (initial RG steps), the main limitation to realizing scale-invariance exactly is physical: the lattice Hamiltonian includes RG irrelevant terms that break scale-invariance at short-distance scales, but are suppressed at larger distances. On the other hand, for large $s$ (large number of RG steps) the main obstruction to scale invariance is the numerical truncation errors, which can be thought of as introducing RG relevant terms, effectively throwing us away from criticality and thus scale invariance. Indeed, use of a larger bond dimension $\chi$, which reduces truncation errors, allows TNR to not only achieve a more precise approximation to scale-invariance, but to hold it for more RG steps.}
\label{fig:ScaleApprox}
\end{center}
\end{figure}

Below the critical temperature, $T = 0.9\ T_C$, we obtain a flow towards a fixed-point tensor $B^{Z_2}$ that has two significant elements $[B^{Z_2}]_{1111}=[B^{Z_2}]_{2222}=0.5$, with all other elements zero or arbitrarily small, corresponding to the ($Z_2$-)symmetry-breaking phase. At criticality, $T = T_C$, the tensors converge to a highly non-trivial fixed point tensor after a small number of RG steps, one which appears only slightly changed from the first tensor $B^{(0)}$. Note that, due to the truncation error of the TNR scheme, we do not obtain a numerically exact fixed point; nonetheless the individual elements of $B^{(2)}$ and $B^{(3)}$ all differ by less than $10^{-4}$, while the largest elements of these tensors are of order $\sim 0.1$. We thus define $B^{\CRI} \equiv B^{(3)}$ as the approximate critical fixed-point tensor. The precision with which scale-invariance is approximated over successive RG steps is further examined in Fig. \ref{fig:ScaleApprox}. Finally, above the critical temperature, $T = 1.1 \ T_C$, we obtain a trivial fixed-point tensor $B^\textrm{triv.}$ that has only a single significant element $[B^{\TRI}]_{1111}=1$ with all other elements zero or arbitrarily small, which is representative of the infinite temperature, disordered phase. 

\section{Appendix B.-- Non-critical RG fixed points: TRG versus TNR.} 

\begin{figure}[!btph]
\begin{center}
\includegraphics[width=6cm]{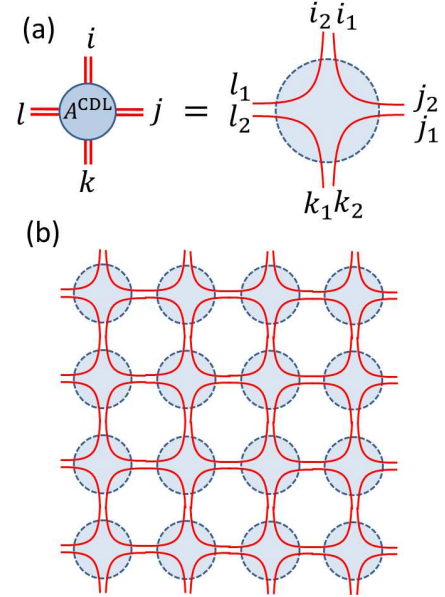}
\caption{(a) CDL tensor $A^{\CDL}$ of Eq. \ref{eq:CDL}. (b) Tensor network made of CDL tensors, which contains correlations only within each plaquette.}
\label{fig:CDL}
\end{center}
\end{figure}

In this appendix we discuss certain aspects of the flow that TRG and TNR generate in the space of tensors, which emphasizes one of the main differences between the two approaches. Specifically, we describe a class of non-critical fixed-points of the flow generated by TRG, namely those represented by corner double line (CDL) tensors $A^{\CDL}$, and show that TNR coarse-grains such tensors into a trivial tensor $A^{\TRI}$. We emphasize that TEFR can also transform a CDL tensor  $A^{\CDL}$ into a trivial tensor $A^{\TRI}$ \cite{TEFR}.

Fig. \ref{fig:CDL}(a) contains a graphical representation of a CDL tensor $A^{\CDL}$, which has components $\left( A^{\CDL} \right)_{ijkl}$ given by
\begin{equation} \label{eq:CDL}
\left( A^{\CDL} \right)_{(i_1 i_2)(j_1 j_2)(k_1 k_2)(l_1 l_2)}\\
= \delta_{i_1j_2} \delta_{j_1k_2} \delta_{k_1 l_2} \delta_{l_1 i_2}
\end{equation}
where a double index notation $i = (i_1 i_2)$ has the meaning $i = i_1 + \eta(i_2 -1)$. Here the double index runs over values $i \in \left\{ 1, 2,\ldots,\eta^2 \right\}$ for integer $\eta$, whereas each single index runs over values $i_1,i_2 \in \left\{1,2,\ldots,\eta \right\}$. Notice that a square network formed from such CDL tensors $A^{\CDL}$ contains only short-ranged correlations; specifically only degrees of freedom within the same plaquette can be correlated. We now proceed to demonstrate that this network is an exact fixed point of coarse-graining with TRG, which was already described in \cite{TEFR}. Note that it is possible to generalize this construction (and the derivation presented below) by replacing each delta in Eq. \ref{eq:CDL} (e.g. $\delta_{i_1j_2}$) with a generic $\eta\times \eta$ matrix (e.g. $M_{i_1 j_2}$) that contains microscopic details. For instance, the fixed point CDL tensors $A^{\CDL} (T)$ obtained with TRG for the off-critical $2D$ classical Ising model partition function would contain matrices $M_{i_1j_2}(T)$ that depend on the initial temperature $T$. However, for simplicity, here we will only consider the case $M_{i_1j_2} = \delta_{i_1j_2}$.

\begin{figure}[!tbph]
\begin{center}
\includegraphics[width=6cm]{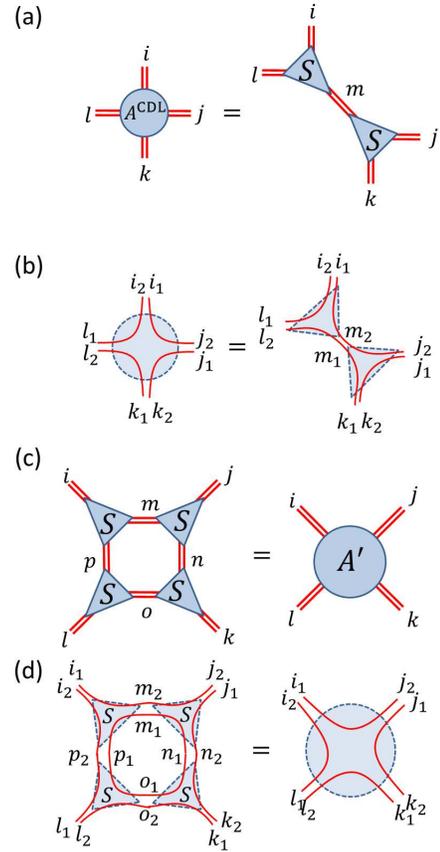}
\caption{A depiction of the two key steps of the TRG coarse-graining transformation. (a) The first step factorizes the tensor $A^{\CDL}$ into a product of two tensors $S$. (b) Local detail of the factorization for CDL tensors, see also Eq. \ref{eq:TRG2}. (c) The second step contracts four tensors $S$ into an effective tensor $A'$. (d) Local detail of the contraction step, see also Eq. \ref{eq:TRG5}. Notice that the effective tensor $A'$ is also a CDL tensor (with a $45$ degree tilt), indicating that CDL tensors are a fixed point of the TRG coarse-graining transformation.}
\label{fig:TRGfixed}
\end{center}
\end{figure}

In the first step of TRG the tensor $A^{\CDL}$ is factorized into a pair of three index tensors $S$,
\begin{equation} \label{eq:TRG2}
\left( A^{\CDL} \right)_{ijkl} = \sum\limits_{m = 1}^{\eta^2} {{S_{mij}}{S_{mkl}}}
\end{equation} 
where, through use of the double index notation introduced in Eq. \ref{eq:TRG2}, tensors $S$ may be written,
\begin{equation} \label{eq:TRG3}
{S_{({m_1}{m_2})({l_1}{l_2})({i_1}{i_2})}} = {\delta _{{m_1}{l_2}}}{\delta _{{l_1}{i_2}}}{\delta _{{i_1}{m_2}}},
\end{equation} 
see also Fig. \ref{fig:TRGfixed}(a-b). The next step of TRG involves contracting four $S$ tensors to form an effective tensor $A'$ for the coarse grained partition function,
\begin{equation} \label{eq:TRG4}
{{A'}_{ijkl}} = \sum\limits_{m,n,o,p = 1}^{\eta^2} {{S_{imp}}{S_{jnm}}{S_{kon}}{S_{lpo}}}
\end{equation} 
where, through use of the explicit form of $S$ given in Eq. \ref{eq:TRG3}, the tensor $A'$ is computed as,
\begin{equation} \label{eq:TRG5}
{{A'}_{({i_1}{i_2})({j_1}{j_2})({k_1}{k_2})({l_1}{l_2})}} = \eta {\delta _{{i_1}{j_2}}}{\delta _{{j_1}{k_2}}}{\delta _{{k_1}{l_2}}}{\delta _{{l_1}{i_2}}},
\end{equation}
see also Fig. \ref{fig:TRGfixed}(c-d). Notice that the effective tensor is proportionate to the original CDL tensor, $A'_{ijkl}=\eta \left( A^{\CDL} \right)_{ijkl}$, where the multiplicative factor of $\eta$ arises from the contraction of a `loop' of correlations down to a point,
\begin{equation} \label{eq:TRG6}
\sum\limits_{{m_1},{n_1},{o_1},{p_1} = 1}^{\eta} {{\delta _{{m_1}{n_1}}}{\delta _{{n_1}{o_1}}}{\delta _{{o_1}{p_1}}}{\delta _{{p_1}{m_1}}}}  = \eta,
\end{equation}
as depicted in Fig. \ref{fig:TRGfixed}(d). That the network of CDL tensors is a fixed point of TRG indicates that some of the short-range correlations in the tensor network are preserved during coarse-graining, i.e. that TRG is artificially promoting short-ranged degrees to a larger length scale. As a result TRG defines a flow in the space of tensors which is not consistent with what is expected of a proper RG flow. Indeed, two tensor networks that only differ in short-range correlations, as encoded in two different $A^{\CDL}$ and $\tilde{A}^{\CDL}$ tensors (differing e.g. in the dimension $\eta$ of each index $i_1, i_2, j_1, \cdots$, or, more generally, on the matrices $M_{i_1 j_2}$ discussed above) should flow to the same fixed point, but under TRG they will not.

\begin{figure}[!tbph]
\begin{center}
\includegraphics[width=5cm]{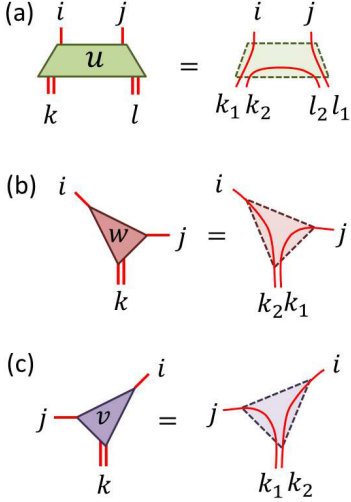}
\caption{Depiction of the explicit form of the required (a) disentangler $u$, see also Eq. \ref{eq:TNR1}, (b-c) isometries $w$ and $v$, see also Eq. \ref{eq:TNR2}, for coarse-graining the network of CDL tensors $A^{\CDL}$, see Eq. \ref{eq:CDL}, using TNR.}
\label{fig:TNRtensors}
\end{center}
\end{figure}

\begin{figure}[!tbph]
\begin{center}
\includegraphics[width=8cm]{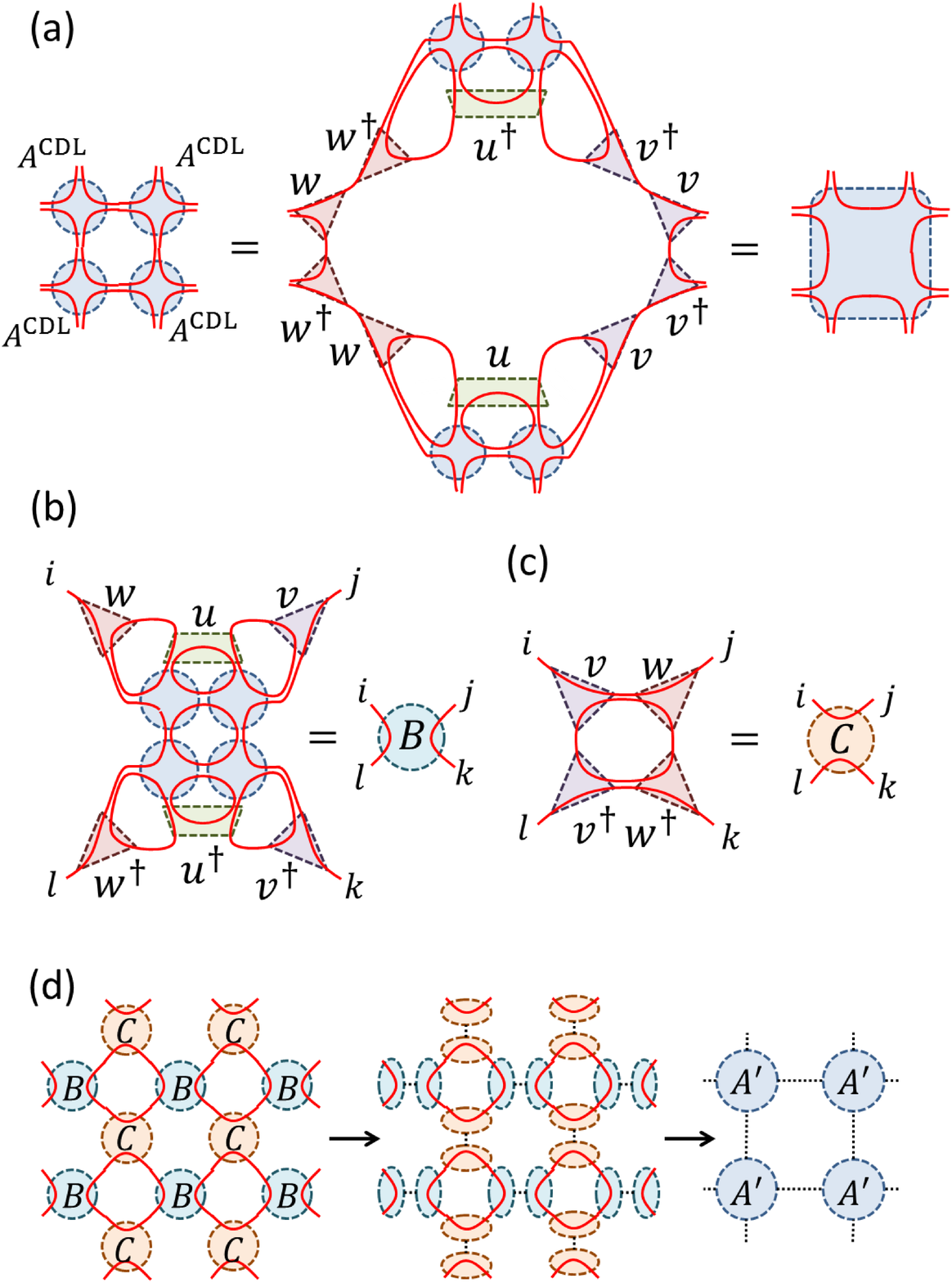}
\caption{A depiction of the key steps of the TNR coarse-graining transformation in the presence of CDL tensors $A^{\CDL}$. (a) Insertion of disentanglers $u$ and isometries $v$ and $w$ between two pairs of CDL tensors $A^{\CDL}$ results in the elimination of the short-range correlations inside the plaquette that those four tensors form. (b)-(c) As a result, the auxiliary tensors $B$ and $C$ only propagate part of the short-range correlations, as represented by the existence of two lines. These tensors are to be compared with the analogous tensor in TRG, namely $A'$ in Fig. \ref{fig:TRGfixed}(d) or $A^{\CDL}$ in Fig. \ref{fig:CDL}(a), which still contain four lines. (d) As a result, the new tensor $A'$, formed by factoring $B$ according to a left-right partition of indices and $C$ according to an up-down partition of indices, is the trivial tensor $A^{\TRI}$, which contains no lines and therefore no correlations}
\label{fig:TNRfixed}
\end{center}
\end{figure}

We next apply the TNR approach to a network of CDL tensors, as defined in Eq. \ref{eq:CDL}, demonstrating that this network is mapped to a network of trivial tensors $A^{\TRI}$, with effective bond dimension $\chi' = 1$, thus further substantiating our claim that the TNR approach can properly address all short-ranged correlations at each RG step. Fig. \ref{fig:TNRtensors} depicts the form of disentangler $u$ and isometries $v$ and $w$ that insert into the network of CDL tensors at the first step of TNR, as per Fig.\ref{fig:TNR}(a),
which are defined as follows. Let us first regard each index of the initial network as hosting a ${\eta}^2$-dimensional complex vector space $\mathbb{V} \equiv \mathbb{C}^{{\eta}^2}$, and similarly define a $\eta$-dimensional complex vector space  $\tilde{\mathbb{V}} \equiv \mathbb{C}^{\eta}$. Then the disentangler $u_{ijkl}$ we use is an isometric mapping $u: \tilde{\mathbb{V}}\otimes \tilde{\mathbb{V}} \rightarrow \mathbb{V} \otimes \mathbb{V}$, with indices $i,j\in\{1,2,\ldots,\eta \}$ and $k,l\in\{1,2,\ldots,{\eta}^2 \}$, that is defined,
\begin{equation} \label{eq:TNR1}
{u_{ij({k_1}{k_2})({l_1}{l_2})}} = \tfrac{1}{{\sqrt \eta }}{\delta _{i{k_1}}}{\delta _{{k_2}{l_2}}}{\delta _{j{l_1}}},
\end{equation}
where we employ the double index notation, $k=(k_1,k_2)$ and $l=(l_1,l_2)$. It is easily verified that $u$ is isometric, $u^{\dagger}u=\tilde{I}^{\otimes 2}$, with $\tilde{I}$ as the identity on $\tilde{\mathbb{V}}$. The isometries $v_{ijk}$ and $w_{ijk}$ are mappings $v: \tilde{\mathbb{V}} \rightarrow \mathbb{V} \otimes \tilde{\mathbb{V}}$ that are defined,
\begin{equation} \label{eq:TNR2}
{w_{ij({k_1}{k_2})}} = {v_{ij({k_1}{k_2})}} = \tfrac{1}{{\sqrt \eta }}{\delta _{i{k_1}}}{\delta _{{k_2}{l_2}}}{\delta _{j{l_1}}},
\end{equation}
where again use double index notation, $k=(k_1,k_2)$. When inserted into the a $2\times 2$ block of tensors from the network of CDL tensors, as depicted in Fig. \ref{fig:uvw}(b),
this choice of unitary $u$ and isometries $w$ and $v$ act as exact resolutions of the identity, see Fig. \ref{fig:TNRfixed}(a). Thus the first step of the coarse graining with TNR, as depicted in Fig. \ref{fig:TNR}(a),
is also exact. Following the second step of TNR, Fig. \ref{fig:TNR}(b),
one obtains four-index tensors $B$ and $C$ as depicted in Fig. \ref{fig:TNRfixed}(b-c), which are computed as,
\begin{align} \label{eq:TNR3}
  {B_{ijkl}} &= {\eta}^4({\delta _{il}}{\delta _{jk}}) \nonumber \hfill, \\
  {C_{ijkl}} &= {\eta}^{-1}({\delta _{ij}}{\delta _{kl}}) \hfill, 
\end{align}
with indices $i,j,k,l \in \{1,2\ldots\eta\}$. The third step of TNR, see Fig. \ref{fig:TNR}(c),
involves factoring the tensors $B$ and $C$ into a products of three index tensors according to a particular partition of indices: $B_{(il)(jk)}$ and $C_{(ij)(kl)}$ respectively. However, we see from Eq. \ref{eq:TNR3} that the tensors factor trivially across these partitions, which implies that the effective tensor $A'$ obtained in the final step of the TNR iteration is trivial (i.e. of effective bond dimension $\chi' = 1$), see also Fig. \ref{fig:TNRfixed}(d). Thus we have confirmed that the network of CDL tensors, which contained only short ranged correlations, can be mapped to a trivial fixed point in a single iteration of TRG, consistent with what is expected of a proper RG flow.

\section{Appendix C.-- Critical systems.}

In this appendix we compare the behaviour of TNR and TRG at criticality. 
TNR produces a critical fixed-point tensor. In contrast, as first argued by Levin and Nave in Ref. \cite{TRG}, TRG does not. Insight into the different behaviour of TRG and TNR is provided by the scaling of an entropy attached to critical tensors, see Fig. \ref{fig:Entropies}. As expected, this entropy is independent of scale in TNR (since TNR produces a critical fixed-point tensor, independent of scale), while it grows roughly linearly with scale in TRG, where no critical fixed-point tensor is reached. 

We also examine improved versions of TRG based on computing a so-called environment, such as the second renormalization group (SRG) \cite{TRGenv,SRG,HOTRG}, and show that they also fail to produce a critical fixed-point tensor -- the entropy grows linearly as in TRG, see Fig. \ref{fig:FlowSRG}. Finally, we point out that in spite of the fact that Ref. \cite{TEFR} refers to TEFR critical tensors as fixed-point tensors, Ref. \cite{TEFR} presents no evidence that TEFR actually produces a fixed-point tensor at criticality.

A critical system is described by a conformal field theory (CFT). We conclude this appendix by reviewing the computation of scaling dimensions of the CFT by diagonalizing a transfer matrix of the critical partition function. Any method capable of accurately coarse-graining the partition function can be used to extract accurate estimates of the scaling dimensions, regardless of whether the method produces a critical fixed-point tensor. Accordingly, Table \ref{table} shows that TRG, an improved TRG with environment, and TEFR indeed produce excellent estimates of the scaling dimensions. Nevertheless, TNR is seen to produce significantly better estimates.

\subsection{Critical fixed-point tensor in TNR}

As discussed in the main text and in appendix A, when applied to the partition function of the critical Ising model, TNR produces a flow $\{A^{(0)}, A^{(1)}, \cdots \}$ in the space of tensors that quickly converges towards a fixed-point tensor $A^{\CRI}$. By fixed-point tensor $A^{\CRI}$ we mean a tensor that remains the same, component by component, under further coarse-graining transformations. This is the case up to small corrections that can be systematically reduced by increasing the bond dimensions $\chi$ (see appendix A for quantitative details).

Thus, at criticality, TNR explicitly recovers scale invariance, as expected of a proper RG transformation. This is analogous to the explicit realization of scale invariance in critical quantum lattice models obtained previously with the MERA \cite{MERA,MERACFT} (and, ultimately, closely related to it \cite{TNRtoMERA}). In addition to making a key conceptual point, realizing scale invariance explicitly in the MERA led to a number of practical results. For instance, it led to identifying a lattice representation of the scaling operators of the theory, and the computation of the scaling dimensions and operator product expansion coefficients of the underlying CFT \cite{MERACFT}. Moreover, it produced an extremely compact representation of the wave-function of a critical system directly in the thermodynamic limit, avoiding finite size effects. In turn, this resulted in a new generation of approaches both for homogeneous critical systems \cite{MERACFT} and for critical systems with impurities, boundaries, or interfaces \cite{MERAboundary}. Finally, the scale-invariant MERA has become a recurrent toy model as a lattice realization of the AdS/CFT correspondence \cite{Swingle}.  We thus expect that the explicit invariance of $A^{\CRI}$, as realized by TNR, will be similarly fruitful.

\subsection{Absence of critical fixed-point tensor in TRG}

In their TRG paper \cite{TRG}, Levin and Nave already argued that TRG should not be expected to produce a fixed-point tensor at criticality. The reason is that a critical tensor in TRG should be thought of as effectively representing a one-dimensional quantum critical system with an amount of entanglement that scales logarithmic in the system size --that is, linearly in the number of coarse-graining step. Thus at criticality the tensor manifestly changes each time that the system is coarse-grained. As a matter of fact, the authors referred to this situation as the breakdown of TRG at criticality (to account for a linear growth of entropy with the number of coarse-graing steps, the bond dimension $\chi$ must grow exponentially, and so the computational cost), and related it to the similar critical breakdown of the density matrix renormalization group \cite{DMRG}. 

In spite of Levin and Nave's original argument justifying the breakdown of TRG at criticality, one might wonder \textit{whether there is a local choice of gauge such that TRG also flows into a fixed-point tensor (up to small numerical errors)} \cite{private}. This local gauge freedom (called field redefinition in \cite{TEFR}), refers to a local change of basis on individual indices of the tensor, implemented by invertible matrices $\mathbf{x}$ and $\mathbf{y}$
\begin{equation}
A_{ijkl} \rightarrow  \sum_{i'j'k'l'} (A)_{i'j'k'l'} \mathbf{x}_{ii'}(\mathbf{x}^{-1})_{k'k} \mathbf{y}_{ll'}(\mathbf{y}^{-1})_{j'j},
\end{equation}
under which the partition function $Z$ represented by the tensor network remains unchanged. That is, one might wonder if after a proper choice of local gauge, the critical tensors $A^{(s)}$ and $A^{(s+1)}$ before and after the TRG coarse-graining step $s$ are (approximately) the same. [We notice that, in order to simplify the comparison with TNR, here by one TRG coarse-graining step we actually mean two TRG coarse-graining steps as originally defined in \cite{TRG}, so that the square lattice is mapped back into a square lattice with the same orientation].

\begin{figure}[!tbph]
\begin{center}
\includegraphics[width=8.5cm]{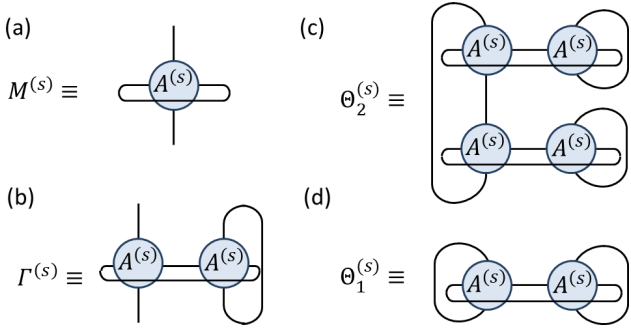}
\caption{
Tensor network representation of matrices $M^{(s)}$ and $\Gamma^{(s)}$, as well as the quantities $\Theta_2^{(s)}$ and $\Theta_1^{(s)}$. By construction, $\Theta_2^{(s)}$ and $\Theta_1^{(s)}$ are invariant under local gauge transformations, as are the eigenvalues of matrices $M^{(s)}$ and $\Gamma^{(s)}$.
}
\label{fig:Tensors}
\end{center}
\end{figure}

In order to definitely confirm that indeed TRG does not produce a critical fixed-point tensor even up to local gauge transformations, we will study the dependence of a matrix $\Gamma^{(s)}$ in the coarse-grainng step $s$. Matrix $\Gamma^{(s)}$ is defined in terms of two tensors $A^{(s)}$ according to Fig. \ref{fig:Tensors}(b). 
$\Gamma^{(s)}$ itself is not invariant under local gauge transformations, but one can build gauge invariant quantities out of it. Here we will focus on two such gauge invariant objects: the spectrum of eigenvalues of $\Gamma^{(s)}$, as well as a simple tensor network $\Theta_2^{(s)}$ built from four copies of $A^{(s)}$ and that amounts to
\begin{equation}
\Theta_2^{(s)} \equiv \tr \left( (\Gamma^{(s)})^2 \right).
\end{equation}

\begin{figure}[!tbph]
\begin{center}
\includegraphics[width=8.5cm]{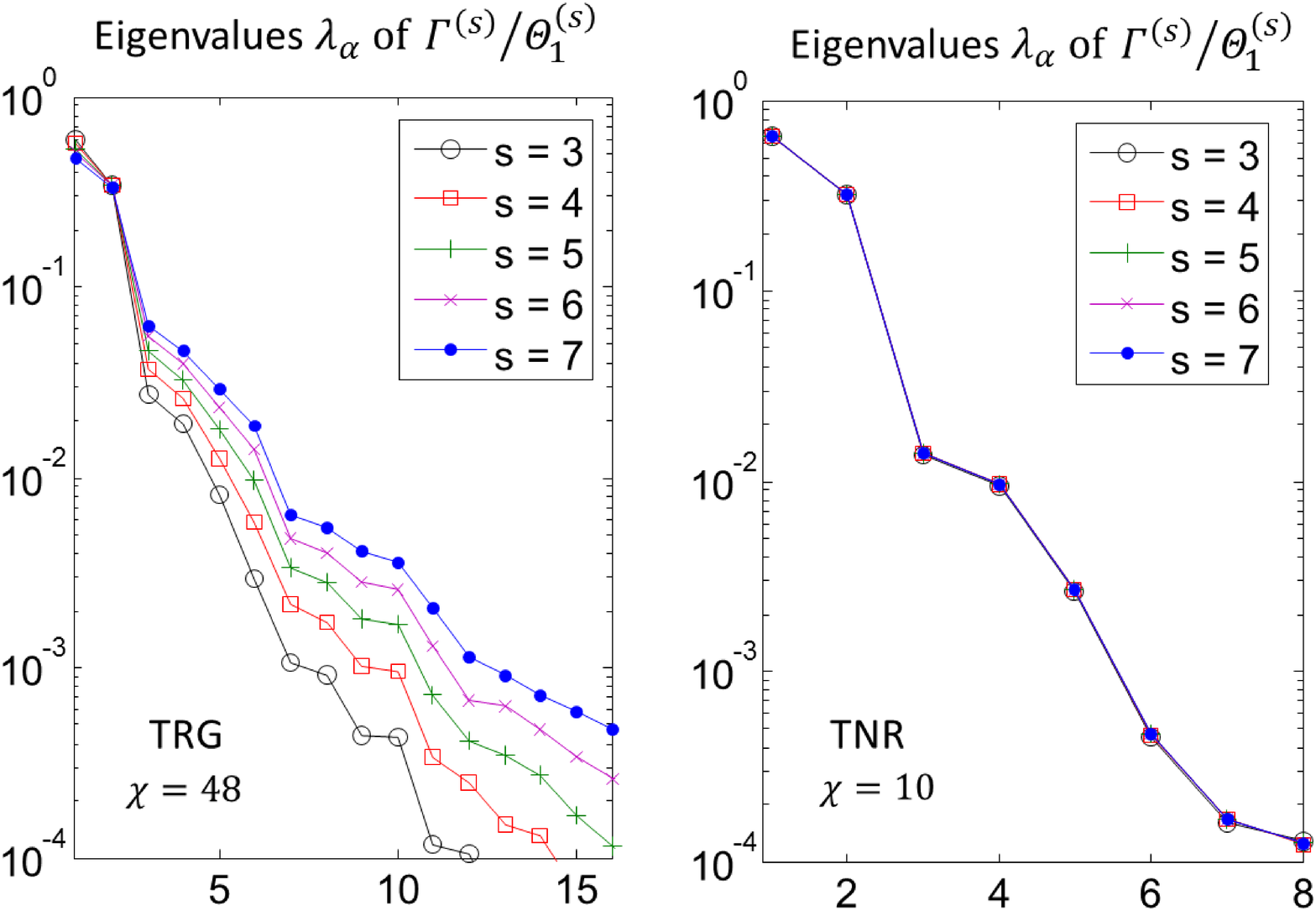}
\caption{
Spectrum of eigenvalues of the normalized matrix $\Gamma^{s}/\Theta_1^{(s)}$ as a function of the coarse-graining step $s$, both for TRG (left) and TNR (right). In TRG, the spectrum becomes flatter with increasing $s$. In TNR, the spectrum is essentially independent of $s$ for $s\geq 3$.
}
\label{fig:Eigenvalues}
\end{center}
\end{figure}

\begin{figure}[!tbph]
\begin{center}
\includegraphics[width=8.5cm]{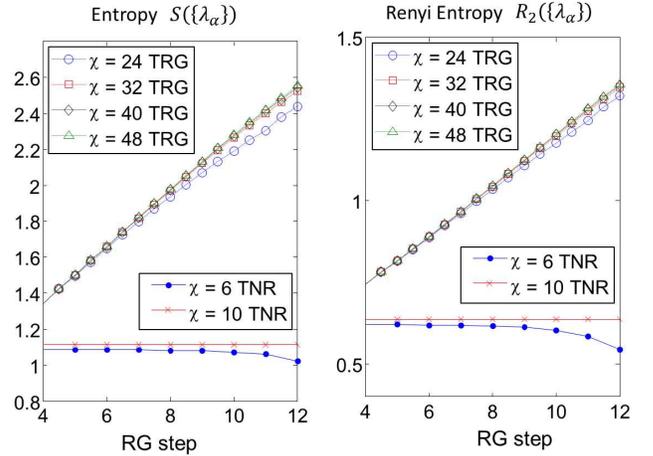}
\caption{
Von Neumman entropy $S(\{\lambda_{\alpha}\})$ and Renyi entropy $R_2(\{\lambda_{\alpha}\})$ of the eigenvalues of the density matrix $\Gamma^{(s)}/\Theta_1^{(s)}$. For TRG, these entropies grow linearly in the coarse-graining step $s$ (or logarithmically in the number $2^{s}$ of quantum spins). For TNR, these entropies are essentially constant for $s\geq 3$.
}
\label{fig:Entropies}
\end{center}
\end{figure}

Fig. \ref{fig:Eigenvalues} shows the spectrum of eigenvalues $\{ \lambda_{\alpha}(s) \}$ of the normalized density matrix $\Gamma^{(s)}/\Theta_1^{(s)}$ as a function of the TRG step $s$. One can see that the spectrum becomes flatter as $s$ increases, indicating that more eigenvalues become of significant size. [In contrast, the spectrum remains essentially constant as a function of the TNR step $s$]. Fig. \ref{fig:Entropies} shows the von Neumannm entropy $S(\{\lambda_{\alpha}\}) \equiv -\sum_{\alpha} \lambda_{\alpha} \log (\lambda_{\alpha})$ and the Renyi entropy $R_n(\{\lambda_{\alpha}\}) \equiv \log (\sum_{\alpha}\lambda_{\alpha}^{n})/(1-n)$ of index $n=2$
\begin{equation}
R_2 = -\log \left( \sum_{\alpha} (\lambda_{\alpha})^2\right) = -\log \left( \frac{\Theta_2^{(s)}}{(\Theta_1^{(s)})^2} \right),
\end{equation}
both of which are invariant under local gauge transformations. These entropies grow linearly in the TRG step $s$. 

The flattening of the spectrum in Fig. \ref{fig:Eigenvalues}, as well as the steady growth of the entropies $S$ and $R_2$ in Fig. \ref{fig:Entropies} constitute an unambiguous confirmation that critical TRG tensors $A^{(s)}$ and $A^{(s+1)}$ are not equal up to a local gauge transformation, not even approximately. In addition, these features are robust against increasing the bond dimension $\chi$, and therefore are not an artifact of the truncation. 

\begin{figure}[!tbph]
\begin{center}
\includegraphics[width=5cm]{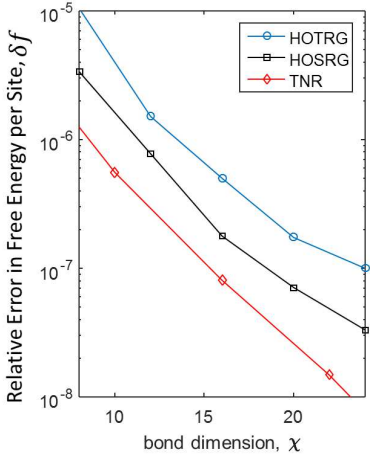}
\caption{Relative error in the free energy per site $\delta f$ as a function of bond dimension $\chi$ for the $2D$ Ising model at criticality, comparing HOTRG, HOSRG and TNR. Through use of the environment HOSRG is able to give better accuracy than HOTRG for any given $\chi$, though the accuracy of HOSRG is still less than that of TNR.}
\label{fig:EnergySRG}
\end{center}
\end{figure}

\subsection{Use of environment}
A significant advance in renormalization methods for tensor networks since the introduction of TRG has been the use of the so-called environment to achieve a more accurate truncation step. The improved TRG method proposed in Ref. \cite{TRGenv} under the name of ``poor-man's SRG" generates local weights that represent the \textit{local} environment in a small neighborhood in the tensor network, while the second renormalization group (SRG) proposed in Ref. \cite{SRG} takes this idea further by using a \textit{global} environment that represents the entir tensor network.


\begin{figure}[!tbph]
\begin{center}
\includegraphics[width=8.5cm]{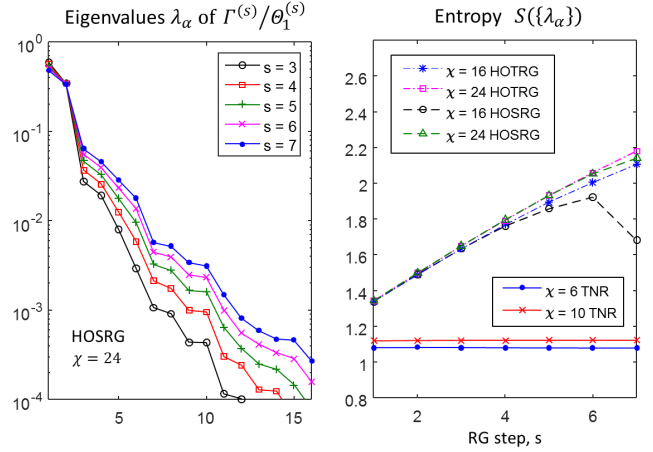}
\caption{ (left) Spectrum of eigenvalues of the normalized matrix $\Gamma^{s}/\Theta_1^{(s)}$ as a function of the coarse-graining step $s$ for HOSRG, In TRG, which becomes flatter with increasing $s$. (right) Von Neumman entropy $S(\{\lambda_{\alpha}\})$ of the eigenvalues of the density matrix $\Gamma^{(s)}/\Theta_1^{(s)}$, comparing HOTRG, HOSRG and TNR. For both HOTRG and HOSRG these entropies grow linearly in the coarse-graining step $s$ (or logarithmically in the number $2^{s}$ of quantum spins), indicating that the tensors are not flowing to a scale-invariant fixed point.}
\label{fig:FlowSRG}
\end{center}
\end{figure}

Fig. \ref{fig:EnergySRG} shows that, for the $2D$ Ising model at criticality, the use of an environment produces a more accurate numerical estimate of the free energy per site. Specifically, the data corresponds to the higher order tensor renormalization group (HOTRG) proposed in Ref.\cite{HOTRG}, an improvement on TRG based upon the higher order singular value decomposition (HOSVD) but does not take the environment into account; and to the higher order second renormalization group (HOSRG), which uses both HOSVD and the global environment. It is also worth pointing out however that TNR, which in its current implementation is a purely local update that does not take into account the environment, produces more accurate results than HOSRG for the same value of the bond dimension. [Moreover, the cost in TNR scales only as $O(\chi^6)$, whereas the cost of HOTRG and HOSRG scales as $O(\chi^7)$].


Based upon the improvement in free energy of HOSRG compared to HOTRG, one may wonder whether the use of the environment, as in SRG, could potentially resolve the breakdown of TRG at criticality and even produce a critical fixed-point tensor. We have investigated this question numerically by analyzing the RG flow of tensors generated by HOTRG and HOSRG for the $2D$ Ising model at criticality, as shown in Fig.\ref{fig:FlowSRG}. Under coarse-graining with HOSRG, the spectrum of the critical tensors grows increasingly flat with RG step, in a similar manner as was observed with standard TRG in Fig.\ref{fig:Eigenvalues}. The same occurs for HOTRG (not plotted). Likewise, in both HOTRG and HOSRG, the entropy of tensors is seen to increase roughly linearly with RG step, and this growth is essentially robust under increasing the bond dimension. These results clearly demonstrate that (HO)SRG does not produce a critical fixed-point tensor. Thus we conclude that the use of environment in (HO)SRG, while leading to a more accurate coarse-graining transformation for fixed bond dimension over (HO)TRG, does not prevent the computational breakdown experienced by TRG at criticality. In particular SRG still fails to produce a proper RG flow and critical fixed-point tensor.

\subsection{The TEFR algorithm}

Ref. \cite{TEFR} refers to the tensors obtained with TEFR at criticality as \textit{fixed-point tensors}. Accordingly, one might be led to conclude that TEFR actually generates a fixed-point tensor (and thus recovers scale invariance) at criticality. As demonstrated in Ref. \cite{TEFR}, TEFR produces accurate estimates of the central charge $c$ and scaling dimensions $\Delta_{\alpha}$ for the Ising model, and one might indeed misinterpret those as being evidence for having generated a critical fixed-point tensor. However, the estimates reported in Ref. \cite{TEFR} for $c$ and $\Delta_{\alpha}$ are of comparable accuracy to those obtained with TRG and ``poor-man's SRG" using the same bond dimension (see Table I in Sect. C.5). Since the above methods demonstrably fail to produce a fixed-point tensor, the accurate TEFR estimates provided in Ref. \cite{TEFR} are no evidence that TEFR has produced a critical fixed-point tensor. In conclusion, Ref. \cite{TEFR} refers to the TEFR critical tensors as fixed-point tensors, but Ref. \cite{TEFR} provides no evidence that TEFR realizes a fixed-point tensor at criticality.

\begin{table}[t]
  \centering
\begin{tabular}{l*{7}{c}r}
            & exact    & TRG(64) & TRG+env(64) & TEFR(64) & TNR(24) \\
\hline
$c$ &	 0.5 	& 0.49982 & 0.49988 & 0.49942 &  0.50001 \\
\hline
$\sigma$& 0.125 & 0.12498 & 0.12498 & 0.12504 &  0.1250004  \\
$\epsilon$&   1 & 1.00055 & 1.00040 & 0.99996 &  1.00009 \\
 & 1.125        & 1.12615 & 1.12659 & 1.12256 &  1.12492  \\
 & 1.125     & 1.12635 & 1.12659 & 1.12403    &  1.12510   \\
 & 2     	& 2.00243 & 2.00549 & - 	& 1.99922 \\ 
 & 2     	& 2.00579 & 2.00557 & - 	& 1.99986 \\ 
 & 2     	& 2.00750 & 2.00566 & - 	& 2.00006 \\ 
 & 2     	& 2.01061 & 2.00567 & - 	& 2.00168  
\end{tabular}
\caption{Exact values and numerical estimates of the central charge $c$ and lowest scaling dimensions of the critical Ising model. TRG results are obtained using the original Levin and Nave's algorithm \cite{TRG}. TRG+env results are obtained using an improved TRG method proposed in Ref. \cite{TRGenv} under the name of ``poor-man's SRG". TEFR results are taken from Ref. \cite{TEFR}. The first three numerical columns use bond dimension $\chi\equiv D_{cut}=64$ and 1024 spins, while the TNR data uses $\chi=24$ and 262,144 spins.}
\label{table}
\end{table}

\subsection{Extraction of scaling dimensions from a transfer matrix}

In proposing TRG in Ref. \cite{TRG}, Levin and Nave refer to the significant loss of efficiency experienced by the method at criticality as \textit{TRG's critical break-down}.  It is important to emphasize that, in spite of this significant loss of efficiency at criticality, TRG can still be used to extract universal information about a phase transition, by studying the partition function of a finite system. 

The key theoretical reason is that in a finite system one can observe a realization of the so-called operator-state correspondence in conformal field theories (CFT), which asserts that there is a one-to-one map between the states of the theory and its scaling operators \cite{CFTbook}. Specifically, the finite size corrections are universal and controlled by the spectrum of scaling dimensions of the theory \cite{Cardy}. This is best known in the context of critical quantum spin chains, where Cardy's formula relates the smallest scaling dimensions $\Delta_{\alpha}$ of the conformal theory to the lowest eigenvalues $E_{\alpha}$ of the critical Hamiltonian on a periodic chain according to
\begin{equation} \label{eq:Cardy}
E_{\alpha}(L) - E_{\alpha}(\infty) = \eta \frac{2\pi}{L} \Delta_{\alpha} + \cdots,
\end{equation}
where $L$ is the number of spins on the chain, $E_{\alpha}(\infty)$ is the energy in an infinite system, $\eta$ (independent of $\alpha$) depends on the normalization of the Hamiltonian, and where the sub-leading non-universal corrections are $O(1/L^2)$ in the absence of marginal operators. Thus, we can estimate scaling dimensions $\Delta_{\alpha}$ by diagonalizing a critical Hamiltonian on a \textit{finite} periodic chain, as it is often done using exact diagonalization techniques. In a two dimensional statistical system, the analogous result is the observation that even in a finite system, a critical partition function $Z$ is organized according to the scaling dimensions of the theory as
\begin{equation} \label{eq:Z1}
Z = e^{a L_xL_y} \sum_{\alpha} e^{-L_y\frac{2\pi}{L_x}(\Delta_{\alpha} - \frac{c}{12}) + \cdots)},
\end{equation}
where $a$ is some non-universal constant, $c$ is the central charge of the CFT, and we assumed isotropic couplings and a square torus made of $L_x \times L_y$ sites. 

\begin{figure}[!tbph]
\begin{center}
\includegraphics[width=7cm]{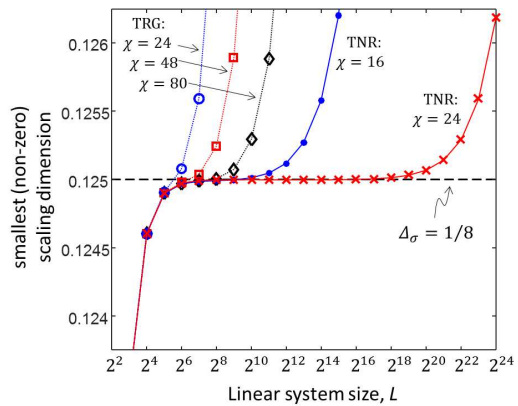}
\caption{Calculation of the smallest non-zero scaling dimension of $2D$ classical Ising at criticality, computed using either TRG or TNR through diagonalization of the transfer operator on an effective linear system size of $L$ spins, comparing to the exact result $\Delta_\sigma = 1/8$. For small system sizes, $L < 2^6$ spins, finite size effects limit the precision with which $\Delta_\sigma$ can be computed, while for large system sizes truncation errors reduce the accuracy of the computed value of $\Delta_\sigma$. However, it is seen that the TNR approach maintains accuracy for much larger system sizes than does TRG: while TRG with bond dimension $\chi=80$ gives $\Delta_\sigma$ within $1\%$ accuracy up to linear system size $L = 2^{11}$ spins, a TNR with bond dimension $\chi = 24$, which required a similar computation time as the $\chi=80$ TRG result, gives $1\%$ accuracy up to $L = 2^{24}$ spins.}
\label{fig:SmallScaleDim}
\end{center}
\end{figure}

As proposed by Gu and Wen \cite{TEFR}, if the coarse-grained tensor $A^{(s)}$ effectively represents the whole partition function of a system made of $2^s \times 2^s$ sites (and thus $L_x=L_y=2^s$), then we can extract the central charge $c$ and scaling dimensions $\Delta_{\alpha}$ from it by e.g. computing the spectrum of eigenvalues of the matrix $M^{(s)}$ (see Fig. \ref{fig:Tensors}(a)), 
\begin{equation}
\left( M^{(s)} \right)_{ik} \equiv \sum_j \left( A^{(s)} \right)_{ijkj}.
\end{equation}
Indeed, after removing the non-universal constant $a$ in Eq. \ref{eq:Z1} [by suitably normalizing the initial tensors $A^{(0)}$], the largest eigenvalues $\lambda_{\alpha}$ of $M^{(s)}$ are of the form \cite{TEFR}
\begin{equation}
\lambda_{\alpha} = e^{-2\pi \left (\Delta_{\alpha} -\frac{c}{12} + O(\frac{1}{L} \right)}.
\end{equation}

For instance, Table \ref{table} shows remarkably accurate estimates of the central charge $c$ and lowest scaling dimensions $\Delta_{\alpha}$ obtained by using TRG and other related methods on a finite lattice made of $1024$ spins. These results are a strong indication that TRG accurately coarse-grains the partition function of such a finite system. [The table also shows more accurate results obtained with TNR on larger systems using a smaller bond dimension.]

In Fig. \ref{fig:SmallScaleDim} we examine the accuracy with which TRG and TNR reproduce the dominant scaling dimension of the Ising model, $\Delta_\sigma = 1/8$, on finite lattices of different linear size $L$. Here it can be seen that TRG results quickly lose accuracy for larger system sizes, a manifestation of the breakdown of TRG at criticality, whilst the results from TNR maintain a high level of accuracy even for very large systems. For instance in comparing the $\chi=80$ TRG result to the $\chi=24$ TNR results, which both required a roughly equivalent computation time to run on a laptop computer ($\sim$ 3 hours), the TRG calculation gave $\Delta_{\sigma}$ to with $1\%$ accuracy for spins systems with linear dimension large as $L=2^{11}$, while the TNR calculation the same accuracy for systems with linear dimension large as $L=2^{24}$: roughly $4000$ times larger linear dimension than TRG.

In conclusion, TRG is an excellent method to study critical partition functions on a finite system, provided that the size of the system that we want to investigate is small enough that the truncation errors are not yet significant. In the critical Ising model, sub-leading (non-universal) finite size corrections to the dominant (universal) contributions to the partition function (that is, to the weights $e^{-2\pi(\Delta_{\alpha} - c/12)}$ for small $\Delta_{\alpha}$) happen to be very small even for relatively small systems. As a result, one can use TRG to extract remarkably accurate estimates of the central charge $c$ and smallest scaling dimensions $\Delta_{\alpha}$. This calculation is analogous to extracting scaling dimensions by diagonalizing the Hamiltonian of a quantum critical system on a finite ring and then using Cardy's formula (Eq. \ref{eq:Cardy}). 

However, the ability to extract accurate estimates of the central charge and scaling dimensions after applying a coarse-graining transformation such as TRG on a finite system does not imply that one has obtained a critical fixed-point tensor. Indeed, we have shown above that TRG does not produce a critical fixed-point tensor, as had already been anticipated by Levin and Nave \cite{TRG}.


\begin{thebibliography}{99}
\bibitem{fixedpoint} In this work (as in Levin and Nave's TRG paper \cite{TRG}) a fixed-point tensor is a tensor that is (explicitly) invariant under coarse-graining \cite{AppendixA}.
\bibitem{npoint} An arbitrary $n$-point correlator will then be computed by first inserting $n$ operators in the tensor network of Eq. \ref{eq:Z}, and then conveniently adjusting its coarse-graining.
\bibitem{disentanglers} The role disentanglers play in TNR, where they remove short-range correlations, is analogous to their role in the context of entanglement renormalization of ground state wave-functions, where they remove short-range entanglement \cite{ER}. Hence the name.
\bibitem{GaugeChange} The gauge transformation is made such that the block of four $A$ tensors is manifestly invariant with respect to reflection along the horizontal axis, which is a convenient choice in order to preserve reflection symmetry in the network.
\bibitem{WeightC} One should take into account the local environment, as described in Ref.\cite{TRGenv}, when decomposing an auxiliary tensor $C^{(s)}$, which may otherwise possess a slowly decaying spectrum of singular values that cannot be accurately truncated.
\bibitem{TEFRalso} For the specific case of CDL tensors, the filtering step of TEFR \cite{TEFR} is also capable of eliminating the local correlations \cite{filterCDL}. TNR removes short-range correlations both away from criticality and near or at a critical point.
\bibitem{fixedboundary} The spontaneous magnetization is extracted from the coarse-grained spin operator, by analysing the projection onto e.g. the spin-up symmetry-breaking sector.
\bibitem{RGstep} We define a single RG step as that which maps the square lattice back into a square lattice with the same orientation, but whose linear dimension has been reduced by a factor of 2. Note that this corresponds to two RG steps as originally proposed in \cite{TRG}.
\end{thebibliography}
\end{document}